\documentclass[fleqn,usenatbib]{mnras}

\usepackage{newtxtext,newtxmath,bm}

\usepackage[T1]{fontenc}

\DeclareRobustCommand{\VAN}[3]{#2}
\let\VANthebibliography\thebibliography
\def\thebibliography{\DeclareRobustCommand{\VAN}[3]{##3}\VANthebibliography}

\usepackage{graphicx}	
\usepackage{amsmath}	

\newcommand{\pderiv}[2]{\frac{\partial#1}{\partial#2}}

\newcommand{\pderivline}[2]{\partial#1/\partial#2}

\newcommand{\sn}[2]{#1\times10^{#2}}

\newcommand{\definealt}{\equiv}

\newcommand{\five}{\ \ \ \ \ }
\newcommand{\orr}{\text{or}\five }
\newcommand{\andd}{\text{and}\five }
\newcommand{\where}{\text{where}\five }

\newcommand{\e}{\hat{\bm{e}}}
\newcommand{\er}{\e_r}

\newcommand{\el}{\e_\lambda}

\newcommand{\rhoref}{\overline{\rho}}
\newcommand{\tmpref}{\overline{T}}
\newcommand{\prsref}{\overline{P}}

\newcommand{\vecg}{\bm{g}}
\newcommand{\geff}{g_{\rm{eff}}}
\newcommand{\vecgeff}{\bm{g}_{\rm{eff}}}

\newcommand{\dsdrline}{d\overline{S}/dr}

\newcommand{\vecu}{\bm{u}}

\newcommand{\rsun}{R_\odot}

\newcommand{\omsun}{\Omega_\odot}

\newcommand{\msun}{M_\odot}
\newcommand{\rstar}{R_*}

\newcommand{\omstar}{\Omega_*}

\newcommand{\rearth}{R_\oplus}
\newcommand{\omearth}{\Omega_\oplus}

\newcommand{\ek}{{\rm{Ek}}}

\newcommand{\ro}{{\rm{Ro}}}
\newcommand{\lo}{{\rm{Lo}}}

\newcommand{\req}{R_{\rm{eq}}}
\newcommand{\rpol}{R_{\rm{pol}}}
\newcommand{\cploc}{c_P}
\newcommand{\hploc}{H_P}
\newcommand{\newtext}[1]{{#1}}
\newcommand{\nnewtext}[1]{{#1}}
\newcommand{\pderivr}{\left(\pderiv{}{\theta}\right)_r}
\newcommand{\pderivp}{\left(\pderiv{}{\theta}\right)_P}

\title[Stellar thermal wind balance]{The stellar thermal wind as a consequence of oblateness}

\author[L. I. Matilsky]{
	Loren I. Matilsky,$^{1}$\thanks{NSF Astronomy and Astrophysics Postdoctoral Fellow}
\\
$^{1}$Department of Applied Mathematics, Baskin School of Engineering, University of California, Santa Cruz\\ \ \ 1156 High St, Santa Cruz, CA 95064, USA; loren.matilsky@gmail.com
}

\date{Accepted XXX. Received YYY; in original form ZZZ}

\pubyear{2023}

\begin{document}
\label{firstpage}
\pagerange{\pageref{firstpage}--\pageref{lastpage}}
\maketitle

\begin{abstract}
\newtext{\nnewtext{In many rotating fluids, the lowest-order force balance is between gravity, pressure, and rotational acceleration (`GPR' balance). Terrestrial GPR balance takes the form of geostrophy and hydrostasy, which together yield the terrestrial thermal wind equation. By contrast, stellar GPR balance is an oblateness equation, which determines the departures of the thermal variables from spherical symmetry; its curl yields the `stellar thermal wind equation.' In this sense, the stellar thermal wind should be viewed not as a consequence of geostrophy, but of baroclinicity in the oblateness.}} Here we treat the \newtext{full stellar oblateness, including the thermal wind, using} pressure coordinates. We derive the generalised stellar thermal wind equation and identify the parameter regime for which it holds. In the case of the Sun, not \nnewtext{considering the full} oblateness has resulted in \newtext{conflicting calculations of the theoretical aspherical temperature anomaly.} We provide new calculation here and find that the \newtext{baroclinic} anomaly is $\sim$3--60 times smaller than the \newtext{barotropic} anomaly. \nnewtext{Thus, the anomaly from the thermal wind} may not be measurable helioseismically; but if measurement were possible, this would potentially yield a new way to bracket the depth of the solar tachocline. 
\end{abstract}




\section{The Terrestrial Thermal Wind}\label{sec:earthtw}
\newtext{In many rotating fluids, the lowest-order force balance is between gravity, pressure, and rotational acceleration (hereafter `GPR' balance). In the Earth's atmosphere and oceans, GPR balance takes the form of hydrostasy in the vertical (i.e., radial) direction (pressure balancing gravity) and geostrophy in the horizontal (i.e., latitudinal and longitudinal) directions (pressure balancing the Coriolis force). When combined, hydrostasy and geostrophy yield the \textit{terrestrial thermal wind equations}. Before briefly describing these equations, we emphasize the special properties of Earth that make it possible to neglect the \nnewtext{fictitious} centrifugal force \nnewtext{(and simultaneously the Earth's oblateness)} from the equations.}

\newtext{First and foremost, \nnewtext{oblateness itself is quite small}: for example, if we define \nnewtext{an object's \textit{geometric oblateness}$f\definealt(\req-\rpol)/\req$ (where $\req$ is the object's equatorial radius and $\rpol$ its polar radius), then $f_\oplus=\sn{3.35}{-3}$ (e.g., \citealt{Chao2006}).} Second, the \nnewtext{shape of the oblateness is such that the Earth's surface is everywhere almost orthogonal to the \textit{effective gravity} $\vecgeff$} (we define $\vecgeff\definealt\vecg_\oplus +\lambda\Omega_\oplus^2\el$, where $\vecg_\oplus$ is \nnewtext{Earth's} gravitational acceleration, $\lambda$ the cylindrical moment arm, \nnewtext{$\Omega_\oplus$ Earth's rotation rate}, and $\e$ denotes a unit vector). The fluid flows can thus be described using spherical coordinates $r,\theta,$ and $\phi$ (radius, colatitude, and longitude, respectively) and $\vecgeff$ becomes purely radial. Ignoring the \nnewtext{fictitious} centrifugal force then leads to a slight error (dependent on latitude) in the magnitude of $\vecgeff$ and ignoring the oblateness leads to geometric errors in, for example, the operators $\pderivline{}{r}$ and $\pderivline{}{\theta}$. Both these errors are of order $f_\oplus$ and do not have any significant dynamical effects.} 

\newtext{One more important property is the thinness of Earth's oceans and atmosphere. The typical \textit{aspect ratio} $\alpha$ of large-scale flow (ratio of the flow's vertical length-scale to its horizontal length-scale) is thus quite small. \nnewtext{Coriolis forces from the horizontal component of angular velocity can then be neglected to order $\alpha$ (e.g., \citealt{Pedlosky1987}, p. 48) and the momentum equation for steady flows splits into a hydrostatic vertical component and a geostrophic horizontal component:}}   
\begin{align}
	\pderiv{P}{r}&=-\rho g_\oplus,\label{eq:hydrostatic}\\
	\andd u_\phi &= \frac{1}{2\omearth\cos\theta} \frac{1}{\rho \rearth}\left(\pderiv{P}{\theta}\right)_r,\label{eq:geostrophic}
\end{align}
\nnewtext{where we have defined the pressure ($P$), density ($\rho$), fluid velocity ($\vecu$), $g_\oplus\definealt|\vecgeff|$ (presumed constant), and the radius of Earth ($\rearth$).} 

\newtext{Throughout this work we explicitly distinguish between colatitudinal derivatives along spherical surfaces (constant $r$, denoted by $(\pderivline{}{\theta})_r$) and isobaric surfaces (constant $P$, denoted by $(\pderivline{}{\theta})_P$).} Given a pressure field $P$, equation \eqref{eq:geostrophic} defines the \textit{geostrophic wind} $u_\phi$. Differentiating equation \eqref{eq:geostrophic} by $r$ and using equation \eqref{eq:hydrostatic} yields the \textit{terrestrial thermal wind equation}:
	\begin{align}\label{eq:thermalwind}
		\pderiv{u_\phi}{r} &=  \frac{1}{2\omearth\cos\theta}\frac{1}{\rho^2} (\nabla P\times\nabla \rho)_\phi\nonumber\\
		&=  -\frac{1}{2\omearth\cos\theta}\frac{g_\oplus}{\rho}  \frac{1}{\rearth} \left(\pderiv{\rho}{\theta}\right)_P.
	\end{align}
\newtext{Equation \eqref{eq:thermalwind} defines the \textit{terrestrial thermal wind} $\pderivline{u_\phi}{r}$---the radial gradient of the geostrophic wind \nnewtext{(e.g., \citealt{Vallis2017}, p. 91)}.} 
 

\nnewtext{Equations \eqref{eq:hydrostatic} and \eqref{eq:geostrophic} also require a small \textit{Rossby number} $\ro$ (ratio of advective force to Coriolis force) and a small \textit{Ekman number} $\ek$ (ratio of viscous force to Coriolis force).} The order to which equation \eqref{eq:thermalwind} holds is thus set by the largest of $f_\oplus$, $\alpha$, $\ro$, and $\ek$. For large-scale atmospheric winds, the rough ordering is $1\gtrsim\ro\gtrsim\alpha\gtrsim f_\oplus\gtrsim\gg\ek$ (e.g., \citealt{Pedlosky1987}).


\section{The Stellar Thermal Wind}
\newtext{In this section, we consider an axisymmetric, differentially rotating star. GPR balance might be expected to yield stellar analogs of the geostrophic and thermal winds. However, the flows in a star do not in general have small aspect ratio, nor are they confined to a surface perpendicular to $\vecgeff$. \nnewtext{It is thus not justifiable to} ignore the \nnewtext{fictitious} centrifugal force, use spherical coordinates while ignoring the oblateness, and assume hydrostatic and geostrophic balance. Nevertheless, this `terrestrial approach' (e.g., \citealt{Thompson2003,Brun2010,Aurnou2011,Matilsky2020b}) is often used to yield a stellar thermal wind equation:}
\begin{align}\label{eq:stellarthermalwindwrong}
	2\Omega_0\pderiv{\omstar}{z}=\frac{\overline{g}}{r^2\sin\theta\overline{\cploc}}\left(\pderiv{S^\prime}{\theta}\right)_r. 
\end{align}
\newtext{Here, $\Omega_*=\Omega_*(\lambda,z)$ is the star's (differential) rotation rate, $\Omega_0$ some typical (constant) `reference' value of $\omstar$, $\cploc$ the specific heat at constant pressure, $S$ the specific entropy, and $z=r\cos\theta$ the axial coordinate. An overbar denotes a spherically-symmetric mean and a prime the deviation from that mean.} 

\newtext{Although approximately correct, equation \eqref{eq:stellarthermalwindwrong} suffers from several issues. First, it is unclear how to choose $\Omega_0$; and in fact, the left-hand side becomes significantly inaccurate for large differential rotation. Second, the colatitudinal derivative on the right-hand side should be at constant $P$, not constant $r$; this makes the equation invalid outside convection zones and obscures how to calculate the other aspherical thermal anomalies (apart from $S^\prime$) due to the thermal wind. Finally, the left-hand side of equation \eqref{eq:stellarthermalwindwrong} comes from the Coriolis force; it should (in the correct equation) come from the \nnewtext{centripetal acceleration}, and thus the Coriolis-based numbers $\text{Ro}$ and $\text{Ek}$ do not define the regime in which stellar thermal wind balance holds. }

\newtext{To derive the correct equation, we work in the inertial frame (although we refer to the centripetal acceleration as a `centrifugal force' when it appears on the right-hand side of the force equation; \nnewtext{this is distinct from the \textit{fictitious} centrifugal force referred to in section \ref{sec:earthtw}}). In the GPR balance equation, we assume that gravity and pressure maintain the centripetal acceleration:}
\begin{subequations}\label{eq:gpr}
	\begin{align}
		-\lambda\Omega_*^2\el  &= -\frac{1}{\rho}\nabla P + \vecg,\label{eq:gpr1}\\
		\orr \nabla P &= \rho\vecgeff, \label{eq:gpr2}\\
		\where \vecgeff&\definealt \vecg + \lambda\omstar^2\el. \label{eq:gpr3}
	\end{align}
\end{subequations}
This equation determines the shape (oblateness) of the star, \nnewtext{or equivalently,} \newtext{the aspherical thermal anomalies $P^\prime$ and $\rho^\prime$ (and thus any other thermal anomaly using the equation of state; e.g., \citealt{Goldreich1968, Armstrong1999})}. It is equivalent to the hydrostatic equation, with the gravity modified by the centrifugal force $\lambda\omstar^2\e_\lambda$. Given fields of $P$, $\rho$, and $\vecg$, equation \eqref{eq:gpr} defines the zonal \textit{asterostrophic wind} $\Omega_*^2$. 

Considering the forces neglected from equation \eqref{eq:gpr}, we define the \textit{GPR Rossby, Lorentz}, and \textit{Ekman numbers}: $\ro_{\rm{GPR}}\definealt U^2/H\lambda \Omega_*^2$ (where $U$ and $H$ are the typical speed and length-scale of flows not associated with the rotation), $\lo_{\rm{GPR}}\definealt B^2/\mu\rho H\lambda \Omega_*^2$ (where $B$ is the strength of the typical magnetic field \newtext{and $\mu$ the magnetic permeability}), and $\ek_{\rm{GPR}}\definealt \nu U/H^2\lambda\Omega_*^2$ (\newtext{where $\nu$ is the kinematic viscosity}). The order to which Equation \eqref{eq:gpr} holds is set by the largest of $\ro_{\rm{GPR}}$, $\lo_{\rm{GPR}}$, and $\ek_{\rm{GPR}}$.



\begin{figure}
	\centering
	\includegraphics[width=3.5in]{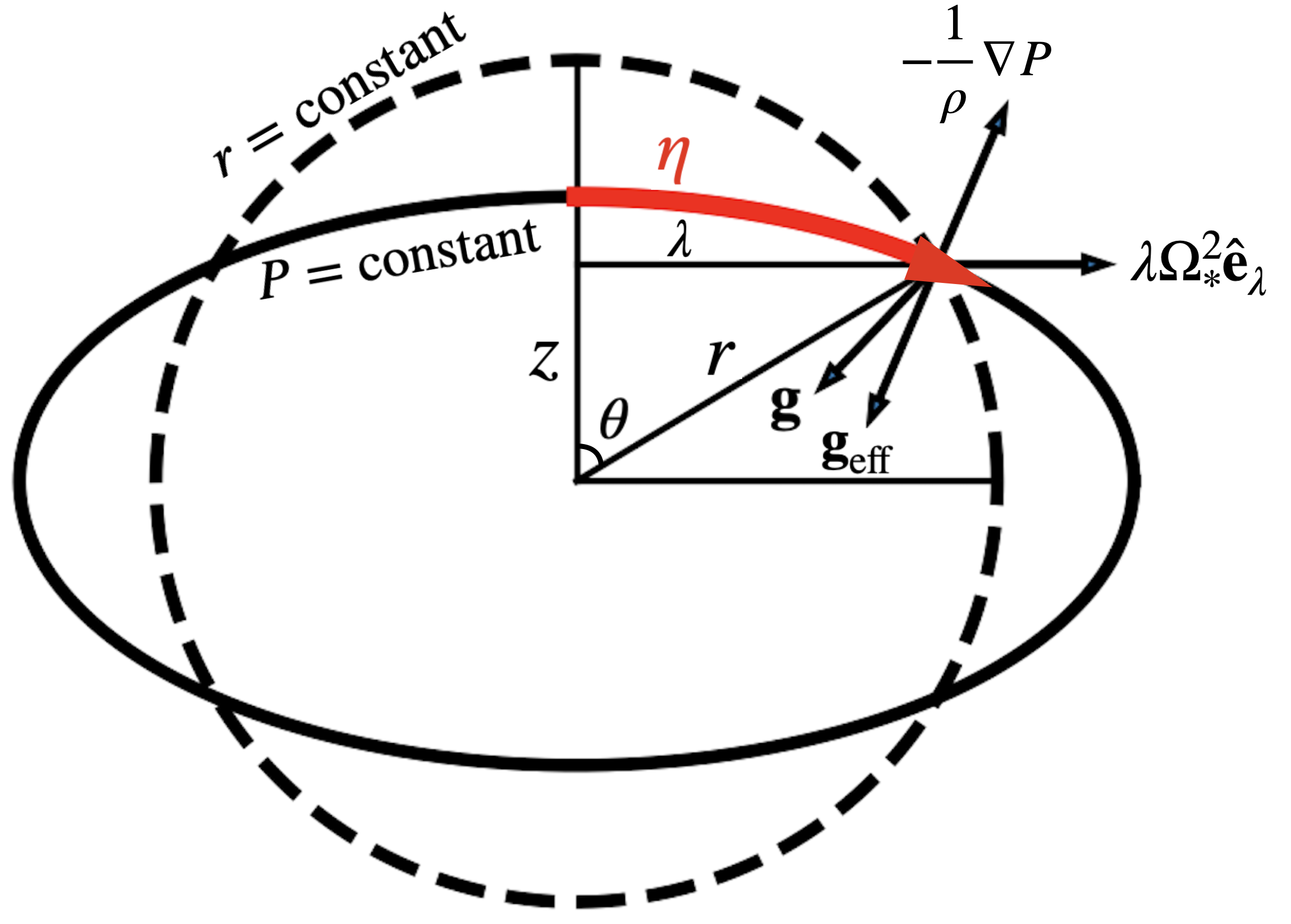}
	\vspace{-0.2in}
	\caption{Schematic showing how the GPR balance equation, \eqref{eq:gpr}, determines the shape of an oblate star by setting the isobaric surfaces. The relation between the different cylindrical, spherical, and pressure coordinates is shown by the labels $r,\theta,\lambda,z,P$, and $\eta$. \nnewtext{Note that $\vecg$ points in the general direction shown (not parallel to $-\er$) because of the oblate centrally concentrated mass.}}
	\label{fig:coord}
\end{figure}

\newtext{Equation \eqref{eq:gpr2} shows that $\vecgeff$ is everywhere parallel $\nabla P$. Thus, the isobaric surfaces determine the natural meanings of `vertical and horizontal', which we define explicitly as `parallel and perpendicular to $\nabla P$', respectively. To describe the thermal wind, we use pressure coordinates. We use $P$ in place of $r$ and} define the colatitude-like coordinate $\eta=\eta(\lambda,z)$ by the distance travelled along an isobaric surface from the Northern rotation axis to $(\lambda,z)$. Figure \ref{fig:coord} schematically shows the GPR force balance and the relationship between the various coordinate systems. 

We take the curl of equation \eqref{eq:gpr} to find, in analogy with equation \eqref{eq:thermalwind}, the equation of \textit{stellar thermal wind balance}:
\begin{align}\label{eq:stellarthermalwind}
	\pderiv{\omstar^2}{z} &= \frac{1}{\lambda}\frac{1}{\rho^2}(\nabla P\times\nabla\rho)_\phi=-\frac{1}{\lambda}\frac{\geff}{\rho}\left(\pderiv{\rho}{\eta}\right)_P,
\end{align}
where $\geff\definealt|\vecgeff|$. We refer to $\pderivline{\omstar^2}{z}$ as the \textit{stellar thermal wind}---the axial variation in the asterostrophic wind. \newtext{This derivation shows that stellar thermal wind balance is simply the baroclinic (curled) component of \nnewtext{the oblateness equation.} The stellar thermal wind \nnewtext{is thus the baroclinic component of the full oblateness}.}

Because the curl has been taken, the conditions under which stellar thermal wind balance holds may be more restrictive than those associated with GPR balance. We define $Z\definealt|\pderivline{\ln\omstar^2}{z}|^{-1}$ as the axial scale of variation of $\omstar^2$ and define the \textit{thermal wind Rossby, Lorentz}, and \textit{Ekman numbers}:
\begin{subequations}\label{eq:twconditions}
\begin{align}
	\ro_{\rm{TW}}    &\definealt \left(\frac{Z}{H}\right)\ro_{\rm{GPR}}=   \left(\frac{Z}{H}\right) \left(\frac{U^2}{H\lambda\omstar^2}\right), \\
	\lo_{\rm{TW}}     &\definealt \left(\frac{Z}{H}\right)\lo_{\rm{GPR}}=   \left(\frac{Z}{H}\right) \left(\frac{B^2}{\mu\rho H \lambda\omstar^2}\right),\\
	\andd \ek_{\rm{TW}}       &\definealt \left(\frac{Z}{H}\right)\ek_{\rm{GPR}} =   \left(\frac{Z}{H}\right) \left(\frac{\nu U}{H^2\lambda\omstar^2}\right).
\end{align} 
\end{subequations}
The order to which Equation \eqref{eq:stellarthermalwind} holds is thus set by the largest of $\ro_{\rm{TW}}$, $\lo_{\rm{TW}}$, and $\ek_{\rm{TW}}$.

We assume local thermodynamic equilibrium (LTE) and a known equation of state to rewrite the right-hand side of equation \eqref{eq:stellarthermalwind} in terms of any other thermodynamic variable. For example, we can instead use the temperature $T$ or the entropy $S$:
\begin{align}\label{eq:eos}
	\frac{1}{\rho}\left(\pderiv{\rho}{\eta}\right)_P = - \beta_T \left(\pderiv{T}{\eta}\right)_P = -\frac{\beta_T T}{\cploc}\left(\pderiv{S}{\eta}\right)_P,
\end{align}
where $\beta_T\definealt-(1/\rho)(\pderivline{\rho}{T})_P$ is the coefficient of expansion.

Note that equations \eqref{eq:stellarthermalwind} and \eqref{eq:eos} make no assumptions about the magnitude of oblateness (i.e., the departures of the thermal variables from spherical symmetry) or the particular equation of state. They are expected to hold anywhere in the star that is in LTE and for which the numbers in equation \eqref{eq:twconditions} are small. \textit{If} the oblateness is small, then $(\pderivline{}{\eta})_P\rightarrow (1/r)(\pderivline{}{\theta})_P$, $\geff\rightarrow g\definealt|\vecg|$, and the thermal variables and $g$ (when not differentiated horizontally) reduce to their spherically symmetric values. Equations \eqref{eq:stellarthermalwind} and \eqref{eq:eos} then become:
\begin{align}\label{eq:reducedthermalwind}
	\pderiv{\omstar^2}{z} &= -\frac{\overline{g}}{r^2\sin\theta}          \frac{1}{\overline{\rho}} \left(\pderiv{\rho^\prime}{\theta}\right)_P =  \frac{\overline{g}}{r^2\sin\theta}     \overline{\beta_T} \left(\pderiv{T^\prime}{\theta}\right)_P\nonumber\\
	 &= \frac{\overline{g}}{r^2\sin\theta}          \frac{\overline{\beta_T}\ \overline{T} }{\overline{\cploc}} \left(\pderiv{S^\prime}{\theta}\right)_P,
\end{align}
which is valid to the order of the numbers in equation \eqref{eq:twconditions} \textit{and} \nnewtext{(say) the geometric} oblateness $f_*$. \newtext{For an ideal gas ($\overline{\beta_T}=1/\overline{T}$) and small differential rotation ($\pderivline{\omstar^2}{z}\approx2\Omega_0\pderivline{\omstar}{z}$), equation \eqref{eq:reducedthermalwind} reduces to equation \eqref{eq:stellarthermalwindwrong}, except that the colatitudinal derivative is now taken along constant $P$ instead of constant $r$. This latter distinction can only be ignored for $S^\prime$ in well-mixed convection zones, since there the radial gradient $\dsdrline$ is the same order as the colatitudinal gradient $(1/r)\pderivline{S^\prime}{\theta}$ (e.g., \citealt{Balbus2012a, Vasil2021}). Prior studies that estimated the solar aspherical thermal anomalies using the thermal wind equation (e.g., \citealt{Matilsky2020b,Choudhuri2021,Vasil2021}) were thus only accurate for $S^\prime$, and only in the convection zone not too close to the photosphere (in the outer $\sim$2\% of the Sun by radius, $\overline{\beta_T}\ \overline{T}/\overline{\cploc}$ ceases to be a constant).}

\newtext{The relative error between the left-hand sides of equations \eqref{eq:stellarthermalwindwrong} and \eqref{eq:stellarthermalwind} (or \eqref{eq:reducedthermalwind}) is equal to the \textit{differential rotation Rossby number} $\ro_{\rm{DR}}\definealt|\omstar-\Omega_0|/\omstar$. Although $\ro_{\rm{DR}}\ll1$ may be true for certain stars, it is not a relevant condition for stellar thermal wind balance. For the Sun in particular, $\ro_{\rm{DR}}$ can be as high as $\sim$0.3, depending on the choice of $\Omega_0$. Apart from increased accuracy, the form $\pderivline{\omstar^2}{z}$ makes it clear that the thermal wind arises not from the Coriolis force (geostrophy), but from the \textit{centrifugal force} (axial variations of which tend to produce zonal vorticity).}
	

The fact that the thermal wind $\pderivline{\omstar^2}{z}$ is a centrifugal term (and should be derived in the inertial frame without introducing the Coriolis force) has been pointed out by numerous authors (e.g., \citealt{Kitchatinov1995,Balbus2009a,Lara2013,Choudhuri2020}). \newtext{We argue there is a deeper point here, namely that in stars, geostrophy simply cannot \nnewtext{hold for} a steady zonal flow. On Earth, whose oblate surface is rigid, applying a latitudinal pressure gradient (`latitudinal' in the strictly horizontal sense, i.e., perpendicular to $\vecgeff$) creates a zonal flow according to geostrophy. In a gaseous star or planet, applying an $\eta$ pressure gradient causes the oblateness to adjust until GPR balance is obtained again. In other words, steadily rotating \nnewtext{axisymmetric} stars do not have nonzero \nnewtext{horizontal} pressure gradients (see Figure \ref{fig:coord}), making geostrophic balance in the \nnewtext{horizontal} direction impossible.}


 \begin{figure*}
	\centering
	\includegraphics[width=7.25in]{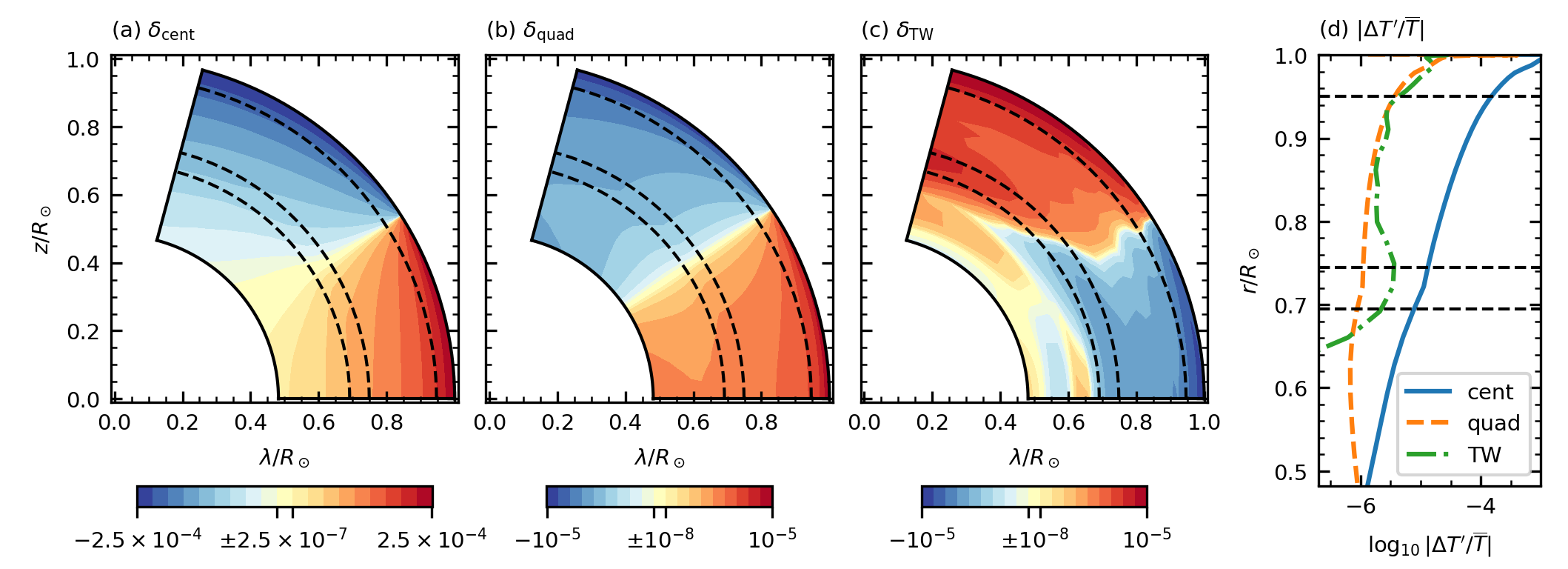}
	\vspace{-0.2in}
	\caption{Contributions to the the solar aspherical temperature anomaly: (a) $\delta_{\rm{cent}}$, (b) $\delta_{\rm{quad}}$, and (c) $\delta_{\rm{TW}}$. They have been calculated from the helioseismic inversion of the rotation rate averaged between 1995--2009 \citep{Howe2005, Howe2023}. Quantities are shown in the upper half of the meridional plane, excluding latitudes greater than $75^\circ$. The colour map is `symmetric logarithmic:' red (blue) tones logarithmically separate positive (negative) values into 9 bins, and one yellow bin contains the values close to zero. The boundaries of each set of bins are given by the colour bar tick labels. Each set of 9 bins spans 3 decades, so neighboring bins are separated by a factor of $\sim$2. (d) Pole-to-equator temperature contrast (|$\Delta\delta|=|\Delta T^\prime/\overline{T}|$) from each contribution. In all panels, \nnewtext{the boundaries of the tachocline and NSSL are marked by dashed black curves}.}
	\label{fig:tmpanomaly}
\end{figure*}


\section{Aspherical Thermal Anomalies}
\newtext{The full aspherical thermal anomalies can only be determined from the oblateness (GPR) equation \eqref{eq:gpr}. By contrast, the curled oblateness equation \eqref{eq:reducedthermalwind} determines just the baroclinic part of the anomaly due to the thermal wind.} We can use the pressure-coordinate formulation to estimate the typical magnitudes of the aspherical thermal anomalies and separate the \newtext{barotropic and baroclinic contributions.} Here we do this for small $f_*$, so $(\pderivline{}{\eta})_P\rightarrow (1/r)(\pderivline{}{\theta})_P$. For the temperature anomaly $T^\prime$, we write \newtext{the purely mathematical statement}:
\begin{align}\label{eq:tmppert}
 \pderivr \left(\frac{T^\prime}{\tmpref}\right)&=\left(\frac{d\ln\tmpref}{d\ln\prsref}\right) \pderivr\left(\frac{P^\prime}{\prsref}\right) + \pderivp \left(\frac{T^\prime}{\tmpref}\right).
\end{align}
On the right-hand side, the first term is from \newtext{the barotropic part of the oblateness and the second term from the baroclinic part (thermal wind). Analogous equations can be written for the other thermal anomalies; note that for $S^\prime$ in a well-mixed convection zone, the barotropic contribution is negligible.}

We write the $\theta$ component of equation \eqref{eq:gpr} as
\begin{align}\label{eq:prspert}
	\pderivr \left(\frac{P^\prime}{\prsref}\right)&=\frac{\rhoref}{\prsref}(r^2\sin\theta\cos\theta\omstar^2 + rg_\theta),
\end{align}
where $g_\theta\definealt\e_\theta\cdot\vecg$. We expect $g_\theta$ to have the same sign and order of magnitude as the centrifugal force term $r\sin\theta\cos\theta\omstar^2$ (Figure \ref{fig:coord}).

To compare the magnitudes of the two terms in equation \eqref{eq:tmppert}, we assume an ideal gas ($\overline{\beta_T} =1/\tmpref$) and hydrostatic balance in the \newtext{spherically symmetric mean variables} ($\overline{g}= \prsref/\rhoref \hploc$, where $\hploc\definealt -(d{\ln \prsref}/dr)^{-1}$ is the local pressure scale-height). Using equations \eqref{eq:reducedthermalwind} and \eqref{eq:prspert}, we estimate:
\begin{subequations}\label{eq:tmppertcompare}
\begin{align}
	\delta_{\rm{bar}}      &\definealt  \left(\frac{d\ln\tmpref}{d\ln\prsref}\right) \int \pderivr \left(\frac{P^\prime}{\prsref}\right)d\theta     \sim \left(\frac{r}{\hploc}\right)      \left(\frac{r\sin\theta\cos\theta\omstar^2}{\overline{g}}\right)\label{eq:pertbar},\\
	\delta_{\rm{TW}}    &\definealt \int \pderivp\left(\frac{T^\prime}{\tmpref}\right)d\theta    \sim  \left(\frac{r}{Z}\right)  \left(\frac{r\sin\theta\omstar^2}{\overline{g}}\right),\label{eq:perttw}
\end{align}
\end{subequations}
\newtext{where we have defined $\delta_{\rm{bar}}$ and $\delta_{\rm{TW}}$ as the temperature anomalies from the barotropic part of the oblateness and the baroclinic part (thermal wind), respectively.}
 Both anomalies are the same magnitude as the oblateness \nnewtext{itself}, since $r\omstar^2/\overline{g} \sim (r/\rstar)^3m$ (where $\rstar$ is the stellar radius and $m\definealt R_*\omstar^2/\overline{g}_{\rm{surf}}$, with  \nnewtext{$\overline{g}_{\rm{surf}}$ the surface gravity) and in general, $m\sim f_*$ (e.g., \citealt{Collins1963})}. However, since  $\delta_{\rm{TW}}/\delta_{\rm{bar}}\sim \hploc/Z$, $\delta_{\rm{TW}}$ is often \newtext{significantly smaller} than $\delta_{\rm{bar}}$, except in regions of very strong axial shear (small $Z$). From equation \eqref{eq:pertbar}, we expect $\delta_{\rm{bar}}$ to be largest near the stellar surface (if the GPR approximation still holds there), since $\hploc$ becomes very small. \newtext{Prior estimates of the solar temperature anomaly---which considered only the thermal wind equation---thus missed the dominant barotropic contribution, as we show explicitly in the following section.}



\section{The solar temperature anomaly}\label{sec:sun}
The Sun's aspherical thermal anomalies (e.g., $T^\prime$ and $S^\prime$) are of fundamental importance. The gradients of $S^\prime$ may determine how efficient the convection is according to mixing-length theory and thus how the radial energy transport varies with latitude (e.g., \citealt{Featherstone2015}). The structure of $T^\prime$ determines the energy transport by the meridional circulation (e.g., \citealt{Matilsky2020b}) and is also closely related to  variations in the sound speed, which may be directly measurable via helioseismology. Observations of the solar emissive flux reveal a slightly superluminous pole \citep{Rast2008}, yielding a pole-to-equator relative difference in emissive flux of $\Delta I/I_0\sim\sn{2}{-3}$, or a difference in effective temperature of $\Delta T_{\rm{eff}}\sim2.5$ K. It is thus natural to speculate that $\Delta T_{\rm{eff}}$ might be an `imprint' of the interior \newtext{temperature anomaly} (e.g., \citealt{Choudhuri2021}).

\begin{table}
	\centering
	\caption{Left two columns: volume-averages of $\Delta T^\prime/\overline{T}$ over the tachocline and the NSSL. Right-most column: volume-average of $\Delta T^\prime$ over the NSSL.}
	\label{tab:dt}
	\begin{tabular}{l *{2}{l} *{1}{r}}	
		\hline
		&	\multicolumn{2}{c}{$\Delta T^\prime/\overline{T}=\Delta\delta$} &	\multicolumn{1}{c}{$\Delta T^\prime$}  \\
		\hline
		&   tachocline & NSSL\ \  & NSSL  \\
		\hline
		centrifugal    & $-\sn{1.1}{-5}$& $-\sn{5.7}{-4}$ & $-39$ K  \\
		quadrupole    & $-\sn{1.0}{-6}$ & $-\sn{1.2}{-5}$ & $-0.85$  K \\
		thermal wind     & $\gtrsim\sn{3.2}{-6}$ & $\sn{1.0}{-5}$ & $1.1$ K \\
		\hline
	\end{tabular}
\end{table}

An obvious way to infer the thermal anomalies is to assume GPR and thermal wind balance and use \newtext{equation \eqref{eq:tmppertcompare}, along with the helioseismically measured solar rotation rate in equations \eqref{eq:reducedthermalwind} and \eqref{eq:prspert}}. However, prior calculations have considered only the thermal wind equation \eqref{eq:reducedthermalwind} (e.g., \citealt{Vasil2021,Jha2021}) and excluded the \newtext{barotropic oblateness contribution}. Furthermore, no distinction was made between $(\pderivline{}{\theta})_P$ and $(\pderivline{}{\theta})_r$, meaning the calculations only yielded $S^\prime$ correctly, and only in the convection zone. Although  \citet{Matilsky2020b} attempted to calculate $T^\prime$ \newtext{(summing both the thermal wind and barotropic parts)}, we used an incorrect form of equation \eqref{eq:prspert}. 

Here we provide a new calculation of the relative temperature anomaly $\delta\definealt T^\prime/\overline{T}$ ($=\delta_{\rm{bar}} + \delta_{\rm{TW}}$), using pressure coordinates. The most uncertainty comes from $g_\theta$, which relates to the distribution of matter in the oblate solar core and has not been measured. To lowest order, we assume that the quadrupole moment dominates and write $g_\theta\approx (3/2)J_2 (G\msun\rsun^2/r^4)\sin\theta\cos\theta$, where $J_2$ is the gravitational quadrupole moment, $G$ the gravitational constant, $\msun$ the solar mass, and $\rsun$ the solar radius. Equation \eqref{eq:prspert} then becomes
\begin{align}\label{eq:prspert2}
	\pderivr \left(\frac{P^\prime}{\prsref}\right)&=\frac{\rhoref}{\prsref}\left[r^2\omstar^2 + \frac{3}{2}(\rsun^2\omsun^2)\left(\frac{J_2}{m_\odot}\right)\left(\frac{\rsun}{r}\right)^3\right]\sin\theta\cos\theta,
\end{align}
where $\omsun\definealt\sn{2.7}{-6}\ \rm{rad\ s^{-1}}$ (or $\omsun/2\pi\definealt430$ nHz) is the \newtext{the approximate solid-body} rotation rate of the radiative zone and $m_\odot\definealt\rsun^3\omsun^2/G\msun=\sn{1.9}{-5}$.

Using equation \eqref{eq:prspert2} \nnewtext{in equation \eqref{eq:pertbar}}, we write $\delta_{\rm{bar}}=\delta_{\rm{cent}} + \delta_{\rm{quad}}$, defining the separate contributions from the centrifugal force and quadrupole. $J_2$ can be estimated from the rotation rate by solving equation \eqref{eq:gpr} simultaneously with the Poisson equation for the gravitational potential (e.g., \citealt{Dicke1967b, Ulrich1981}). Using the helioseismic rotation law then yields $J_2\approx\sn{2.2}{-7}$ \citep{Pijpers1998,Armstrong1999}. \newtext{Note that $(3/2)J_2/m_\odot\approx0.017$, so that for the Sun, $\delta_{\rm{quad}}\ll \delta_{\rm{cent}}$, except deep in the interior.}

We take the spherically averaged quantities from model S \citep{ChristensenDalsgaard1996}. We use equations \eqref{eq:reducedthermalwind} and \eqref{eq:prspert2} in \eqref{eq:tmppertcompare} to find $\delta_{\rm{cent}}$, $\delta_{\rm{quad}}$, and $\delta_{\rm{TW}}$. We set the integration constants so that each $\delta_{(\cdots)}$ separately has no spherical mean (excluding the region above $75^\circ$ latitude). For $T^\prime=T^\prime(r,\theta)$, we also define the pole-to-equator temperature contrast: $\Delta T^\prime(r) \definealt T^\prime(r,\pi/12) - T^\prime(r,\pi/2)$ \nnewtext{and $\Delta\delta=\Delta T^\prime/\overline{T}$} (positive for \nnewtext{for a pole hotter than the equator}). The result is shown in figure \ref{fig:tmpanomaly} and table \ref{tab:dt}. \nnewtext{In presenting these results, we refer to the Sun's two radial shear layers: the tachocline (radial boundaries at $(0.72\pm0.025)\rsun$) and the near-surface shear layer (NSSL; boundaries at $0.95\rsun$, $\rsun$).}


The biggest contribution is $\delta_{\rm{cent}}$ (figure \ref{fig:tmpanomaly}a). This term would make the equatorial regions slightly hotter than the polar regions ($\Delta T^\prime<0$), with $|\Delta T^\prime/\overline{T}|$ increasing from $\sim$$10^{-5}$ in the tachocline to $\sim$$\sn{6}{-4}$ in the NSSL (table \ref{tab:dt}). Helioseismic measurements so far have been sensitive to asphericity on the order of $\sim$$10^{-4}$ in the relative sound speed anomaly (e.g., \citealt{Antia2001}), which should match $\delta$ where the gas is ideal. Thus, $\delta_{\rm{cent}}$ (and not $\delta_{\rm{TW}}$, which is $\sim$$10^{-5}$ in the NSSL) is the only contribution currently measurable by helioseismology, and only in the outermost layers (e.g., the NSSL). \newtext{To further complicate a possible future measurement of thermal wind balance, $\delta_{\rm{TW}}$ and $\delta_{\rm{quad}}$ appear to be nearly equal and opposite above \nnewtext{$r\approx0.8\rsun$}.} Unless $J_2$ is known accurately, subtracting \nnewtext{an estimated} $\delta_{\rm{quad}}$ from a measured $\delta$ could uncover false thermal wind balance. 

Interestingly, \citet{Antia2001} measure a sound speed in the NSSL that is \textit{smaller} near the equator than near the poles (i.e., $\Delta T^\prime>0$). This implies that possibly GPR balance does not hold in the NSSL, or (as mentioned by \citealt{Antia2001}), the measurements reflect the interior magnetic field, not the sound speed variations.
 
Deep in the Sun, the helioseismically measured $Z$ is really an upper bound, set by the large size of the inversion kernels (e.g., \citealt{Howe2009}, p. 31). This makes \nnewtext{our estimated value of} $\delta_{\rm{TW}}$ a lower bound. It has been proposed that the tachocline is substantially thinner than the inversion kernels (e.g., \citealt{Elliott1997}), \nnewtext{but the true thickness $\Gamma$ is still unknown and would have important consequences for the global solar dynamo and the tachocline confinement problem (e.g., \citealt{Miesch2005,Matilsky2022}). If we accept thermal wind balance a priori, then a precisely measured $\delta$ in the tachocline (after subtracting the $\delta_{\rm{cent}}$ and $\delta_{\rm{quad}}$ estimated here) would yield the true value of $\delta_{\rm{TW}}$, and hence $\Gamma$. For example, the inversion considered here \citep{Howe2005,Howe2023} gives $\Gamma\lesssim0.05\rsun$ and $\Delta\delta_{\rm{TW}}\gtrsim\sn{3}{-6}$ (table \ref{tab:dt}). If a future measurement of the sound speed anomaly were to yield $\Delta\delta\approx10^{-5}$ ($\Delta\delta_{\rm{TW}}\approx\sn{2}{-5}$), that would imply $\Gamma\approx0.01\rsun$.} 

We point out that none of the values in table \ref{tab:dt} (either dimensional or non-dimensional) closely match the $\Delta I/I_0\sim\sn{2}{-3}$ or $\Delta T_{\rm{eff}}\sim 2.5$ K reported in \citet{Rast2008}. \nnewtext{Even ignoring the discrepancies in sign, the magnitudes of} our non-dimensional values are all too low (as are the dimensional $|\Delta T^\prime|$ from the thermal wind and quadrupole), while $|\Delta T^\prime|$ from the centrifugal force is too high. This indicates that the interior temperature anomaly does \textit{not} imprint directly onto the surface. This may not be too surprising \newtext{(and does necessarily mean GPR or thermal wind balance \nnewtext{breaks down in} the NSSL)}, since it is still not well-understood how the interior $T^\prime$ and $S^\prime$ determine the energy flux (and thus the observed $\Delta T_{\rm{eff}}$, which really measures asphericity in the outward energy flux at the photosphere). 

\newtext{Apart from measuring the thermal anomalies, the validity of solar GPR and thermal wind balance could be assessed by estimating the relevant non-dimensional numbers in equation \eqref{eq:twconditions} from the \nnewtext{expected} Reynolds and Maxwell stresses. In the NSSL, for example, supergranulation gives $U\sim100\ \rm m\ s^{-1}$ and $H\sim30$ Mm \nnewtext{(e.g., \citealt{Rincon2018})}. Using $\lambda\sim500$ Mm, $\omstar\sim\sn{3}{-6}\ \rm rad\ s^{-1}$, and $Z\sim10^3$ Mm \nnewtext{(for $Z$, see \citealt{Matilsky2023b})} gives $\ro_{\rm{GPR}}\sim0.07$ and $\ro_{\rm{TW}}\sim2.5$. We thus might expect the NSSL to be in GPR balance, but not thermal wind balance (ignoring for now the Maxwell stress since its value is \nnewtext{poorly} constrained). Of course, this estimate implicitly assumes that the meridional components of the velocity from supergranulation are perfectly correlated; a poor correlation might yield $\ro_{\rm{TW}}\ll1$, \nnewtext{and then thermal wind balance might still hold}.}

\section{Conclusions}
We derive the equations of GPR balance and the corollary of \textit{stellar} thermal wind balance, which we contrast with the different equation of \textit{terrestrial} thermal wind balance. We also derive the non-dimensional numbers that must be small to make \nnewtext{each equation} valid. We show that the stellar thermal wind is really a consequence of the centrifugal force (oblateness) and not of the Coriolis force (geostrophy). Furthermore, the aspherical thermal anomalies due to the \newtext{barotropic} parts of the oblateness (here we consider \nnewtext{the contributions from} the centrifugal force and \nnewtext{gravitational} quadrupole moment) are necessarily the same order of magnitude as the \newtext{baroclinic} anomaly from the thermal wind, and often substantially larger.

Using pressure coordinates, we estimate the temperature anomaly for the Sun, separating the contributions from the centrifugal force, gravitational quadrupole, and thermal wind. We find that verifying or falsifying thermal wind balance by measuring the solar temperature anomaly heloseismically may not be possible. On the other hand, measuring the anomaly to an improved precision of $\sim$$10^{-5}$ (and simply accepting thermal wind balance for the deep interior) may yield a new way to bracket the width of the tachocline. Finally, it is not clear exactly how measurements of surface emissive flux (e.g., \citealt{Rast2008}) connect to the interior thermal anomalies. Until this connection is well-understood (\newtext{or the interior Reynolds and Maxwell stresses are more reliably estimated}), it is still an open question whether GPR and thermal wind balance hold \nnewtext{everywhere in the Sun}. 

\section*{Acknowledgements}
I was supported during this work by a National Science Foundation Astronomy \& Astrophysics Postdoctoral Fellowship under award AST-2202253. I thank Nicholas H. Brummell, Bradley W. Hindman, Pascale Garaud, Juri Toomre, Jonathan M. Aurnou, and Terry A. Matilsky for helpful discussions, as well as the anonymous reviewer for extremely helpful feedback. I am indebted to Rachel Howe for providing invaluable helioseismic inversion data and I thank the Daviau-Naidu Institute for their hospitality and encouragement. 

\section*{Data Availability}
The helioseismic inversion of the solar rotation rate \nnewtext{is publicly available via Zenodo \citep{Howe2023}}. Model S is available at \href{https://users-phys.au.dk/jcd/solar_models/}{https://users-phys.au.dk/\textasciitilde jcd/solar\_models/}. The solar temperature anomalies, \nnewtext{entropy anomalies, and $Z$ (with a more complete description of numerical methods)} are publicly \nnewtext{available via Zenodo \citep{Matilsky2023b}}. 
 




\begin{thebibliography}{}
	\makeatletter
	\relax
	\def\mn@urlcharsother{\let\do\@makeother \do\$\do\&\do\#\do\^\do\_\do\%\do\~}
	\def\mn@doi{\begingroup\mn@urlcharsother \@ifnextchar [ {\mn@doi@}
		{\mn@doi@[]}}
	\def\mn@doi@[#1]#2{\def\@tempa{#1}\ifx\@tempa\@empty \href
		{http://dx.doi.org/#2} {doi:#2}\else \href {http://dx.doi.org/#2} {#1}\fi
		\endgroup}
	\def\mn@eprint#1#2{\mn@eprint@#1:#2::\@nil}
	\def\mn@eprint@arXiv#1{\href {http://arxiv.org/abs/#1} {{\tt arXiv:#1}}}
	\def\mn@eprint@dblp#1{\href {http://dblp.uni-trier.de/rec/bibtex/#1.xml}
		{dblp:#1}}
	\def\mn@eprint@#1:#2:#3:#4\@nil{\def\@tempa {#1}\def\@tempb {#2}\def\@tempc
		{#3}\ifx \@tempc \@empty \let \@tempc \@tempb \let \@tempb \@tempa \fi \ifx
		\@tempb \@empty \def\@tempb {arXiv}\fi \@ifundefined
		{mn@eprint@\@tempb}{\@tempb:\@tempc}{\expandafter \expandafter \csname
			mn@eprint@\@tempb\endcsname \expandafter{\@tempc}}}
	
	\bibitem[\protect\citeauthoryear{Antia \& Basu}{Antia \&
		Basu}{2001}]{Antia2001}
	Antia H.~M.,  Basu S.,  2001, \mn@doi [ApJ] {10.1086/323701}, 559, L67
	
	\bibitem[\protect\citeauthoryear{Armstrong \& Kuhn}{Armstrong \&
		Kuhn}{1999}]{Armstrong1999}
	Armstrong J.,  Kuhn J.~R.,  1999, \mn@doi [ApJ] {10.1086/307879}, 525, 533
	
	\bibitem[\protect\citeauthoryear{Aurnou \& Aubert}{Aurnou \&
		Aubert}{2011}]{Aurnou2011}
	Aurnou J.~M.,  Aubert J.,  2011, \mn@doi [PEPI] {10.1016/j.pepi.2011.05.011},
	187, 353
	
	\bibitem[\protect\citeauthoryear{Balbus}{Balbus}{2009}]{Balbus2009a}
	Balbus S.~A.,  2009, \mn@doi [MNRAS] {10.1111/j.1365-2966.2009.14469.x}, 395,
	2056
	
	\bibitem[\protect\citeauthoryear{Balbus, Latter  \& Weiss}{Balbus
		et~al.}{2012}]{Balbus2012a}
	Balbus S.~A.,  Latter H.,   Weiss N.,  2012, \mn@doi [MNRAS]
	{10.1111/j.1365-2966.2011.20217.x}, 420, 2457
	
	\bibitem[\protect\citeauthoryear{Brun, Antia  \& Chitre}{Brun
		et~al.}{2010}]{Brun2010}
	Brun A.~S.,  Antia H.~M.,   Chitre S.~M.,  2010, \mn@doi [A\&A]
	{10.1051/0004-6361/200913166}, 510, A33
	
	\bibitem[\protect\citeauthoryear{Chao}{Chao}{2006}]{Chao2006}
	Chao B.~F.,  2006, \mn@doi [CRGeo] {10.1016/j.crte.2006.09.014}, 338, 1123
	
	\bibitem[\protect\citeauthoryear{Choudhuri}{Choudhuri}{2020}]{Choudhuri2020}
	Choudhuri A.~R.,  2020, \mn@doi [SCPMA] {10.1007/s11433-020-1628-1}, 64
	
	\bibitem[\protect\citeauthoryear{Choudhuri}{Choudhuri}{2021}]{Choudhuri2021}
	Choudhuri A.~R.,  2021, \mn@doi [SoPh] {10.1007/s11207-021-01784-7}, 296
	
	\bibitem[\protect\citeauthoryear{Christensen-Dalsgaard
		et~al.,}{Christensen-Dalsgaard et~al.}{1996}]{ChristensenDalsgaard1996}
	Christensen-Dalsgaard J.,  et~al., 1996, \mn@doi [Sci.]
	{10.1126/science.272.5266.1286}, 272, 1286
	
	\bibitem[\protect\citeauthoryear{Collins}{Collins}{1963}]{Collins1963}
	Collins G.~W.,  1963, \mn@doi [ApJ] {10.1086/147712}, 138, 1134
	
	\bibitem[\protect\citeauthoryear{Dicke}{Dicke}{1967}]{Dicke1967b}
	Dicke R.~H.,  1967, \mn@doi [ApJ] {10.1086/180072}, 149, L121
	
	\bibitem[\protect\citeauthoryear{{Elliott}}{{Elliott}}{1997}]{Elliott1997}
	{Elliott} J.~R.,  1997, A\&A, \href
	{https://ui.adsabs.harvard.edu/abs/1997A&A...327.1222E} {327, 1222}
	
	\bibitem[\protect\citeauthoryear{Featherstone \& Miesch}{Featherstone \&
		Miesch}{2015}]{Featherstone2015}
	Featherstone N.~A.,  Miesch M.~S.,  2015, \mn@doi [ApJ]
	{10.1088/0004-637x/804/1/67}, 804, 67
	
	\bibitem[\protect\citeauthoryear{Goldreich \& Schubert}{Goldreich \&
		Schubert}{1968}]{Goldreich1968}
	Goldreich P.,  Schubert G.,  1968, \mn@doi [ApJ] {10.1086/149821}, 154, 1005
	
	\bibitem[\protect\citeauthoryear{Howe}{Howe}{2009}]{Howe2009}
	Howe R.,  2009, \mn@doi [LRSP] {10.12942/lrsp-2009-1}, 6, 1
	
	\bibitem[\protect\citeauthoryear{Howe}{Howe}{2023}]{Howe2023}
	Howe R.,  2023, Mean solar rotation profile from GONG splittings 1995-2009,
	\mn@doi{10.5281/ZENODO.8171572}
	
	\bibitem[\protect\citeauthoryear{Howe, Christensen-Dalsgaard, Hill, Komm, Schou
		\& Thompson}{Howe et~al.}{2005}]{Howe2005}
	Howe R.,  Christensen-Dalsgaard J.,  Hill F.,  Komm R.,  Schou J.,   Thompson
	M.~J.,  2005, \mn@doi [ApJ] {10.1086/497107}, 634, 1405
	
	\bibitem[\protect\citeauthoryear{Jha \& Choudhuri}{Jha \&
		Choudhuri}{2021}]{Jha2021}
	Jha B.~K.,  Choudhuri A.~R.,  2021, \mn@doi [MNRAS] {10.1093/mnras/stab1717},
	506, 2189
	
	\bibitem[\protect\citeauthoryear{{Kitchatinov} \& {Ruediger}}{{Kitchatinov} \&
		{Ruediger}}{1995}]{Kitchatinov1995}
	{Kitchatinov} L.~L.,  {Ruediger} G.,  1995, A\&A, \href
	{https://ui.adsabs.harvard.edu/abs/1995A&A...299..446K} {299, 446}
	
	\bibitem[\protect\citeauthoryear{Lara \& Rieutord}{Lara \&
		Rieutord}{2013}]{Lara2013}
	Lara F.~E.,  Rieutord M.,  2013, \mn@doi [A\&A] {10.1051/0004-6361/201220844},
	552, A35
	
	\bibitem[\protect\citeauthoryear{Matilsky}{Matilsky}{2023}]{Matilsky2023b}
	Matilsky L.~I.,  2023, Dataset for MNRAS Letter: The stellar thermal wind as a
	consequence of oblateness, \mn@doi{10.5281/ZENODO.8176830}
	
	\bibitem[\protect\citeauthoryear{Matilsky, Hindman  \& Toomre}{Matilsky
		et~al.}{2020}]{Matilsky2020b}
	Matilsky L.~I.,  Hindman B.~W.,   Toomre J.,  2020, \mn@doi [ApJ]
	{10.3847/1538-4357/ab9ca0}, 898, 111
	
	\bibitem[\protect\citeauthoryear{Matilsky, Hindman, Featherstone, Blume  \&
		Toomre}{Matilsky et~al.}{2022}]{Matilsky2022}
	Matilsky L.~I.,  Hindman B.~W.,  Featherstone N.~A.,  Blume C.~C.,   Toomre J.,
	2022, \mn@doi [ApJL] {10.3847/2041-8213/ac93ef}, 940, L50
	
	\bibitem[\protect\citeauthoryear{Miesch}{Miesch}{2005}]{Miesch2005}
	Miesch M.~S.,  2005, \mn@doi [LRSP] {10.12942/lrsp-2005-1}, 2, 1
	
	\bibitem[\protect\citeauthoryear{Pedlosky}{Pedlosky}{1987}]{Pedlosky1987}
	Pedlosky J.,  1987, Geophysical Fluid Dynamics.
	Springer, New York, \mn@doi{10.1007/978-1-4612-4650-3}
	
	\bibitem[\protect\citeauthoryear{Pijpers}{Pijpers}{1998}]{Pijpers1998}
	Pijpers F.~P.,  1998, \mn@doi [MNRAS] {10.1046/j.1365-8711.1998.01801.x}, 297,
	L76
	
	\bibitem[\protect\citeauthoryear{Rast, Ortiz  \& Meisner}{Rast
		et~al.}{2008}]{Rast2008}
	Rast M.~P.,  Ortiz A.,   Meisner R.~W.,  2008, \mn@doi [ApJ] {10.1086/524655},
	673, 1209
	
	\bibitem[\protect\citeauthoryear{Rincon \& Rieutord}{Rincon \&
		Rieutord}{2018}]{Rincon2018}
	Rincon F.,  Rieutord M.,  2018, \mn@doi [LRSP] {10.1007/s41116-018-0013-5}, 15,
	6
	
	\bibitem[\protect\citeauthoryear{Thompson, Christensen-Dalsgaard, Miesch  \&
		Toomre}{Thompson et~al.}{2003}]{Thompson2003}
	Thompson M.~J.,  Christensen-Dalsgaard J.,  Miesch M.~S.,   Toomre J.,  2003,
	\mn@doi [ARA\&A] {10.1146/annurev.astro.41.011802.094848}, 41, 599
	
	\bibitem[\protect\citeauthoryear{Ulrich \& Hawkins}{Ulrich \&
		Hawkins}{1981}]{Ulrich1981}
	Ulrich R.~K.,  Hawkins G.~W.,  1981, \mn@doi [ApJ] {10.1086/158992}, 246, 985
	
	\bibitem[\protect\citeauthoryear{Vallis}{Vallis}{2017}]{Vallis2017}
	Vallis G.~K.,  2017, Atmospheric and Oceanic Fluid Dynamics.
	Cambridge University Press, \mn@doi{10.1017/9781107588417}
	
	\bibitem[\protect\citeauthoryear{Vasil, Julien  \& Featherstone}{Vasil
		et~al.}{2021}]{Vasil2021}
	Vasil G.~M.,  Julien K.,   Featherstone N.~A.,  2021, \mn@doi [PNAS]
	{10.1073/pnas.2022518118}, 118
	
	\makeatother
\end{thebibliography}








\bsp	
\label{lastpage}
\end{document}